\documentclass{article}

\usepackage[english]{babel}

\usepackage{tabularx}
\usepackage[letterpaper,top=2cm,bottom=2cm,left=3cm,right=3cm,marginparwidth=1.75cm]{geometry}

\def\BibTeX{{\rm B\kern-.05em{\sc i\kern-.025em b}\kern-.08em
    T\kern-.1667em\lower.7ex\hbox{E}\kern-.125emX}}
    
\usepackage{cite}

\usepackage{amsmath}
\usepackage{graphicx}
\usepackage[colorlinks=true, allcolors=blue]{hyperref}

\title{Circuit-theoretic Line Outage Distribution Factor}
\author{Shimiao~Li, Amritanshu~Pandey, Larry~Pileggi \\
Carnegie Mellon University\\
\{shimiaol, amritanp, pileggi\}@andrew.cmu.edu }

\begin{document}
\date{}

\maketitle
\begin{abstract}
    This work presents the design of AC line outage distribution factor created from the circuit-theoretic power flow models. Experiment results are shown to demonstrate its efficacy in quantifying the impact of line outages on the grid, and its resulting potential for fast contingency screening.
\end{abstract}

\section{Introduction}
Sensitivity analysis approximates the impact of specific grid contingencies on AC system without having to run the the nonlinear power flow simulations. It is able to safeguard the power grid and enhance the operator's efficiency by making fast predictions from linearized system models. 

However, today's commonly-used distribution factors\cite{lodf}\cite{generalized-lodf}\cite{gdf}\cite{ptdf}\cite{df-variants} are mainly based on DC power flow assumptions, which are a set of unrealistic constraints making the linear approximation far from the real-world AC system. Moreover, most traditional distribution factors, e.g., line outage distribution factor (LODF)\cite{lodf}\cite{generalized-lodf} only quantify the impact of disturbance in terms of real power, but do not look into other important dimensions, like the impact on current magnitude of transmission lines. These limitations will prevent the sensitivities from giving accurate and reliable predictions. 

In recent years, the circuit-theoretic approaches for power flow analysis\cite{sugar-pf} have made possible an expressive linear system representation at the current operating point, under rectangular coordinate. Here we present the design of AC sensitivity factors created from the circuit-theoretic power flow models. We take the line outage sensitivity as the example to illustrate our idea. Experiment results are shown to demonstrate its efficacy in quantifying the impact of line outages on the grid, and its resulting potential for fast contingency screening\cite{nultiple-contingency-screening}. 

\section{More background: AC vs DC sensitivity factors}
An accurate description and analysis of the power grid requires building the AC power flow model characterized by full AC power flow equations% built from information of generation, load and the relevant network. 
%$$AC equations$$
, obtaining the solution to which needs iterative Newton-Raphson updates. Upon convergence, the last iteration provides a linear approximation of the AC system around the present operating point:
\begin{equation}
    \begin{bmatrix}
\Delta P\\ \Delta Q
\end{bmatrix} =   
    \begin{bmatrix}
\frac{\partial P}{\partial \theta} & \frac{\partial P}{\partial |V|}\\
\frac{\partial Q}{\partial \theta} & \frac{\partial Q}{\partial |V|}
\end{bmatrix}
\begin{bmatrix}
\Delta \theta\\ \Delta |V|
\end{bmatrix}
=J\begin{bmatrix}
\Delta \theta\\ \Delta |V|
\end{bmatrix}
    \label{Ac power flow model}
\end{equation}
where $|V|$ denotes a vector of voltage magnitude, $\theta$ denotes a vector of phase angle, the $J$ is called Jacobian matrix. 

For sake of an efficient problem-solving, an
approximate decoupling of active and reactive flows is enabled by assuming 1p.u. voltage, all angles being close, and lossless line. These assumptions (usually called DC power flow assumptions) simplify the Jacobian matrix and reduces the original model to the decoupled power flow model. Further ignoring the balance of reactive power gives rise to the DC power flow model
\begin{equation}
    \Delta P=B\Delta\theta
    \label{DC power flow model}
\end{equation}
where $B$ is the imaginary part of the admittance matrix with shunts removed. 

This simple linear approximation of power grid allows for fast grid analysis under conditions of limited computation resource, especially for grid control centers several decades ago. And in existing grid analytical toolboxes (e.g. powerworld, matpower, etc), distribution factors are commonly obtained from the DC power flow model, to estimate the linear impact of grid changes or contingencies.

\section{Traditional line outage distribution factor(LODF)}
Given an outage on a certain line $l$, line outage distribution factor (LODF)\cite{lodf}\cite{generalized-lodf} quantifies how much the outage affects real power flow on any monitored line $m$ in the systemm through the ratio between power change $\Delta P_{m}$ on the monitored line and the pre-outage real power on the outage line $l$:
\begin{align*}
LODF_{m,l}=\frac{\Delta P_{m}}{P^{pre-outage}_{l}}
\end{align*}

To enable calculation from power flow model, the line outage is modelled by a transfer between terminals of the outage line, %which is further equivalent to the superposition of power injections at the two terminal nodes in equal amount but opposite direction, 
as Figure \ref{fig:LODF} shows. When the outage line is closed, setting up a power transfer (or injections) of the post-injection line flow
%enforces the rest of the grid to have no power interaction with this line, and thus
makes it equivalent to disconnecting the line physically. The value of the post-injection line flow is unknown before injection and needs to be calculated.
\begin{figure}[h]
	\centering
	\includegraphics[width=0.2\linewidth]{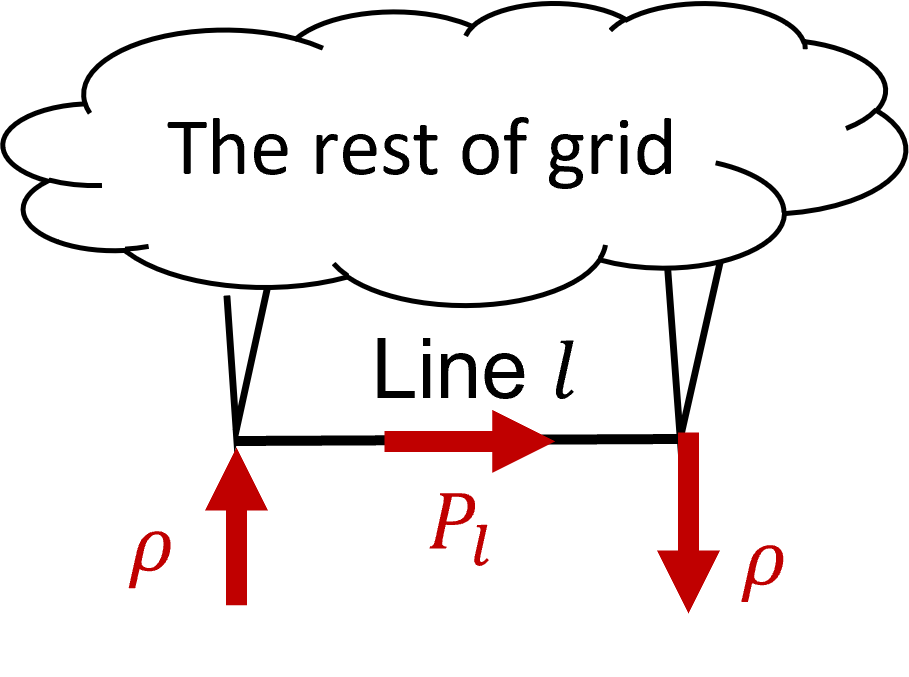}
	\caption[]{Setting up power injections at two terminals of the line $l$ (two injections of the same amount $\rho$ but opposite directions) to model line outage. Let note the line power flow after injection, if $\rho=P_l$, then line $l$ is equivalent to being physically disconnected.
}
	\label{fig:LODF}
\end{figure}

A similar factor, namely line closure distribution factor (LCDF), examines the closure of a presently open line.
Under conditions of multiple line outages or closures, generalized LODF and LODF factors have been proposed to capture the joint impact.

\section{Circuit-theoretic line outage sensitivity}
Taking line outage as an example, our circuit-theoretic sensitivity analysis quantifies the impact of grid changes through a 3-step process:
\begin{enumerate}
    \item Obtain a circuit-theoretic linear system approximation
    \item Model outage event by current injection
    \item Calculate target sensitivity factors
\end{enumerate}
\subsection{Linear circuit-theoretic model of power system}

As an interconnection of facilities that generate, deliver and consume electricity, power grid has a circuit nature. Instead of describing components with their $P,Q,|V|, \delta$ parameters, the circuit theoretic framework models each component within the grid as an equivalent circuit by its current-voltage (I-V) relationship under the rectangular coordinate. These relationships can represent both transmission and distribution grids without loss of generality and, for computational analyticity, can be split into real and imaginary parts, resulting in real and imaginary sub-circuits whose nodes correspond to power system buses. A linear parameterization of the circuit model gives rise to a circuit-theoretic linear AC system model characterized by a set of linear Kirchhoff's Current Law (KCL) equations:
\begin{equation}
    I=YV
    \label{eq:linear cicuit model}
\end{equation}
where $I=[I^{real}_1,I^{imag}_1,...,I^{real}_n,I^{imag}_n]^T$ is the vector of real and imaginary bus injection current, $V=[V^{real}_1,V^{imag}_1,...,V^{real}_n,V^{imag}_n]^T$ is the state vector of real and imaginary voltage, for a system with $n$ buses.

As the circuit-theoretic power flow simulation goes, a nonlinear set of system equations is solved iteratively until convergence. This nonlinearity of power flow model comes from the nonlinear nature of generator and loads. Upon convergence, the last iteration provides a linear approximation of the system, in the form of eq. (\ref{eq:linear cicuit model}). 
\subsection{Current injection to model a line outage}
According to Substitution theorem: under conditions of unique circuit solution, one can replace any element with an independent voltage/current source that constrains the same value of its voltage/current and does not change the rest of the circuit. This indicates an easy way to model the existence, addition, and deletion of elements.

\begin{figure}[h]
	\centering
	\label{fig: sugar line outage}
	\includegraphics[width=0.55\linewidth]{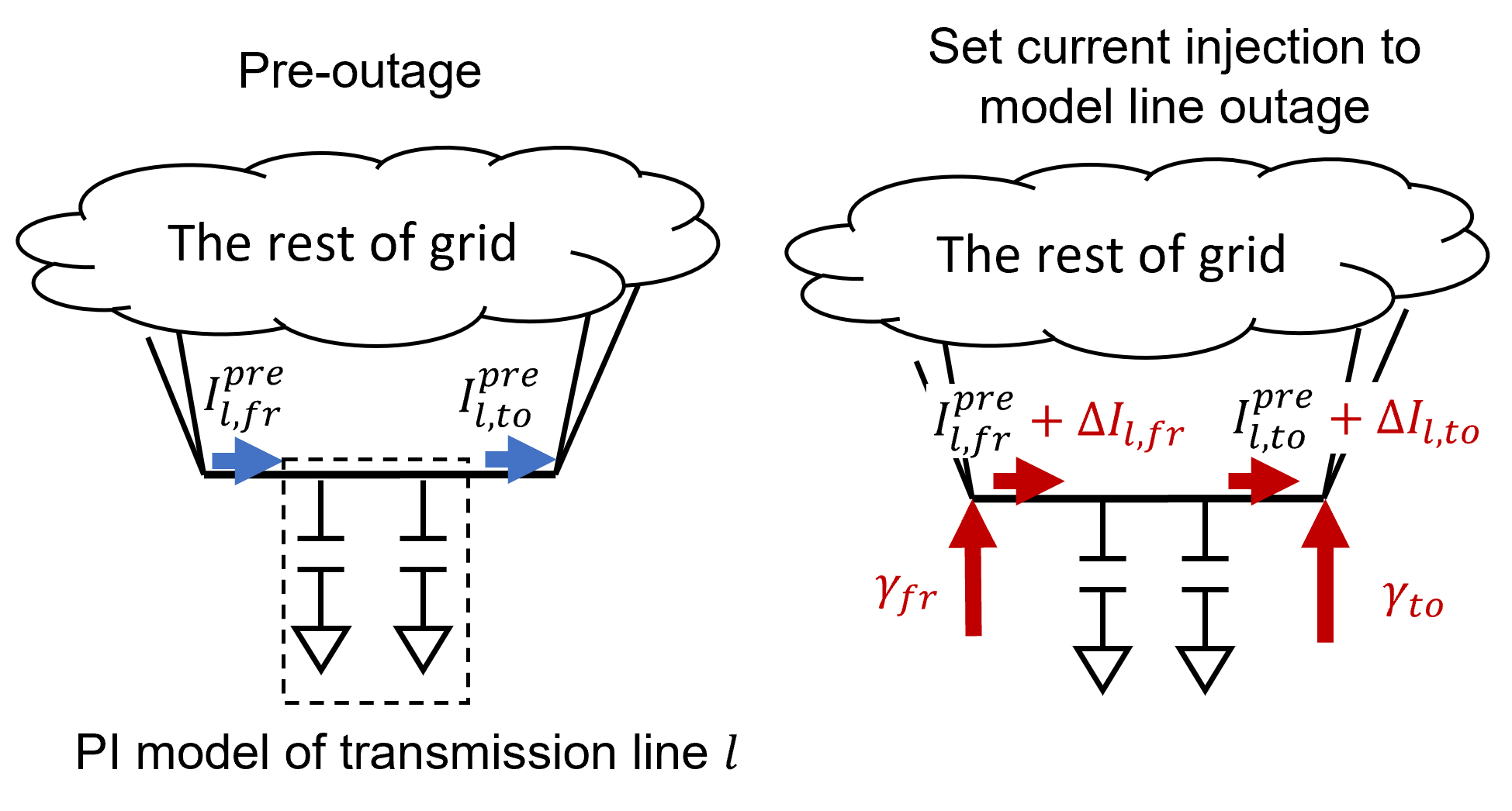}
	\caption[]{Setting up current injections $\gamma$ to model line outage. If line $I_l^{pre}+\Delta I_l=\gamma$ at each end, then line $l$ is equivalent to being physically disconnected.
}
	\label{fig:LODF}
\end{figure}

Line outage factors play a prominent rule in analysing cascading outages.
In the circuit-theoretic model, any line outage can be modelled by injecting current at the two ends of the outage line on the pre-outage system. As Figure \ref{fig: sugar line outage} shows, given an outage line $l$ and a monitored line $m$, line $l$ can be treated equivalently as physically disconnected from the grid, by injecting $\gamma$ at two terminals $i$ and $j$ to make 
\begin{equation}
    \begin{bmatrix}
 I_{l,fr}^{r,pre}\\  I_{l,fr}^{i,pre} \\
  I_{l,to}^{r,pre}\\  I_{l,to}^{i,pre}
\end{bmatrix} +   
    \begin{bmatrix}
 \Delta I_{l,fr}^{r}\\ 
 \Delta I_{l,fr}^{i} \\
 \Delta  I_{l,to}^{r}\\  
 \Delta I_{l,to}^{i}
\end{bmatrix}
=
\begin{bmatrix}
\gamma_{fr}^r\\
\gamma_{fr}^i\\
\gamma_{to}^r\\
\gamma_{to}^i
\end{bmatrix}
    \label{eq: line outage and injection}
\end{equation}
where $r/i$ denotes real/imaginary, $I_{l,fr/to}^{r/i,pre}$ denotes the pre-outage current on the from/to end of line $l$; $\Delta I_{l,fr/to}$ denotes the current change on line $l$ caused by injection $\gamma$.

%After such injection, line $l$ can be treated equivalently as physically disconnected from the grid. The injection makes changes to the rest of grid in the same way as what an actually line outage would cause. 
Yet the value of $\gamma$ is unknown. How can we get it? 

From the system model $I-YV$, the linear impact of any current injection $\frac{dV}{d\gamma}$ is obtainable from $Y^{-1}$. Further making use of chain rule and network parameters, the sensitivity of branch current w.r.t current injection can be calculated 
\begin{equation}
    \frac{dI_{line}}{d\gamma}=\frac{dI_{line}}{dV}\frac{dV}{d\gamma}
    \label{dIline/dgamma}
\end{equation}

Then by approximating $\Delta I_{l,fr/to}=\frac{dI_{line}}{d\gamma}\gamma$, and substituting it into (\ref{eq: line outage and injection}), we have
\begin{equation}
    \begin{bmatrix}
 I_{l,fr}^{r,pre}\\  I_{l,fr}^{i,pre} \\
  I_{l,to}^{r,pre}\\  I_{l,to}^{i,pre}
\end{bmatrix} +   
    \begin{bmatrix}
\frac{dI_{l,fr}^r}{d\gamma_{fr}^r} &
\frac{dI_{l,fr}^r}{d\gamma_{fr}^i} &
\frac{dI_{l,fr}^r}{d\gamma_{to}^r} &
\frac{dI_{l,fr}^r}{d\gamma_{to}^r}\\
\frac{dI_{l,fr}^i}{d\gamma_{fr}^r} &
\frac{dI_{l,fr}^i}{d\gamma_{fr}^i} &
\frac{dI_{l,fr}^i}{d\gamma_{to}^r} &
\frac{dI_{l,fr}^i}{d\gamma_{to}^r}\\
\frac{dI_{l,to}^r}{d\gamma_{fr}^r} &
\frac{dI_{l,to}^r}{d\gamma_{fr}^i} &
\frac{dI_{l,to}^r}{d\gamma_{to}^r} &
\frac{dI_{l,to}^r}{d\gamma_{to}^r}\\
\frac{dI_{l,to}^i}{d\gamma_{fr}^r} &
\frac{dI_{l,to}^i}{d\gamma_{fr}^i} &
\frac{dI_{l,to}^i}{d\gamma_{to}^r} &
\frac{dI_{l,to}^i}{d\gamma_{to}^r}
\end{bmatrix}
\begin{bmatrix}
\gamma_{fr}^r\\
\gamma_{fr}^i\\
\gamma_{to}^r\\
\gamma_{to}^i
\end{bmatrix}
=
\begin{bmatrix}
\gamma_{fr}^r\\
\gamma_{fr}^i\\
\gamma_{to}^r\\
\gamma_{to}^i
\end{bmatrix}
    \label{eq: line outage}
\end{equation}
i.e.,
\begin{equation}
    \begin{bmatrix}
 I_{l,fr}^{r,pre}\\  I_{l,fr}^{i,pre} \\
  I_{l,to}^{r,pre}\\  I_{l,to}^{i,pre}
\end{bmatrix} =   
T
\begin{bmatrix}
\gamma_{fr}^r\\
\gamma_{fr}^i\\
\gamma_{to}^r\\
\gamma_{to}^i
\end{bmatrix}
    \label{eq: line outage final}
\end{equation}
where $T=\begin{bmatrix}
1-\frac{dI_{l,fr}^r}{d\gamma_{fr}^r} &
-\frac{dI_{l,fr}^r}{d\gamma_{fr}^i} &
-\frac{dI_{l,fr}^r}{d\gamma_{to}^r} &
-\frac{dI_{l,fr}^r}{d\gamma_{to}^r}\\
-\frac{dI_{l,fr}^i}{d\gamma_{fr}^r} &
1-\frac{dI_{l,fr}^i}{d\gamma_{fr}^i} &
-\frac{dI_{l,fr}^i}{d\gamma_{to}^r} &
-\frac{dI_{l,fr}^i}{d\gamma_{to}^r}\\
-\frac{dI_{l,to}^r}{d\gamma_{fr}^r} &
-\frac{dI_{l,to}^r}{d\gamma_{fr}^i} &
1-\frac{dI_{l,to}^r}{d\gamma_{to}^r} &
-\frac{dI_{l,to}^r}{d\gamma_{to}^r}\\
-\frac{dI_{l,to}^i}{d\gamma_{fr}^r} &
-\frac{dI_{l,to}^i}{d\gamma_{fr}^i} &
-\frac{dI_{l,to}^i}{d\gamma_{to}^r} &
1-\frac{dI_{l,to}^i}{d\gamma_{to}^r}
\end{bmatrix}$

With $I_{l,r/to}^{r/i,pre}$ obtainable from the pre-outage system model, the solution to (\ref{eq: line outage final}) gives the value of injection $\gamma$. 

Till now we are able to model the line outage by injecting current $gamma$ on the pre-contingency system.

\subsection{Diversified sensitivity factors available from chain rule}

With the outage modelled by a known current injection, the line outage impact $\Delta V$ can be calculated easily from:
\begin{equation}
    \Delta V=
\begin{bmatrix}
\frac{dV}{d\gamma_{fr}^r} &
\frac{dV}{d\gamma_{fr}^i} &
\frac{dV}{d\gamma_{to}^r} &
\frac{dV}{d\gamma_{to}^r}
\end{bmatrix}
       T^{-1} \begin{bmatrix}
 I_{l,fr}^{r,pre}\\  I_{l,fr}^{i,pre} \\
  I_{l,to}^{r,pre}\\  I_{l,to}^{i,pre}
\end{bmatrix}
\end{equation}

Upon availability of the change on state variables $\Delta V$ caused by any disturbance and contingency, the linear impact on diversified dimensions can also be quantified through use of chain rule. For example, given any disturbance/outage, its linear impact on line current magnitude can be estimated through
\begin{equation}
    \Delta |I_{line}|=\frac{\Delta |I_{line}|}{\Delta V}\Delta V
\end{equation}
where $\frac{\Delta |I_{line}|}{\Delta V}$ is obtainable from network parameters; and $\Delta V$ is available from the linear impact quantified above.

These diversified sensitivities go beyond quantifying the impact on power redistribution, but provides a number of other assessment dimensions of interest, including line thermal limit using $\Delta |I_{line}|$, voltage drop limit using $\Delta |V_{line}|$, and stability limit for long-distance transmission using $\Delta P_{line}$.

\section{Results}
One potential application of sensitivity analysis is contingency screening\cite{nultiple-contingency-screening}. Given a large amount of possible outages, we can filter out the top $K$ dangerous contingency events from their sensitivity quantification.
\begin{figure}[h]
	\centering
	\label{fig:case2383}
	\includegraphics[width=0.7\linewidth]{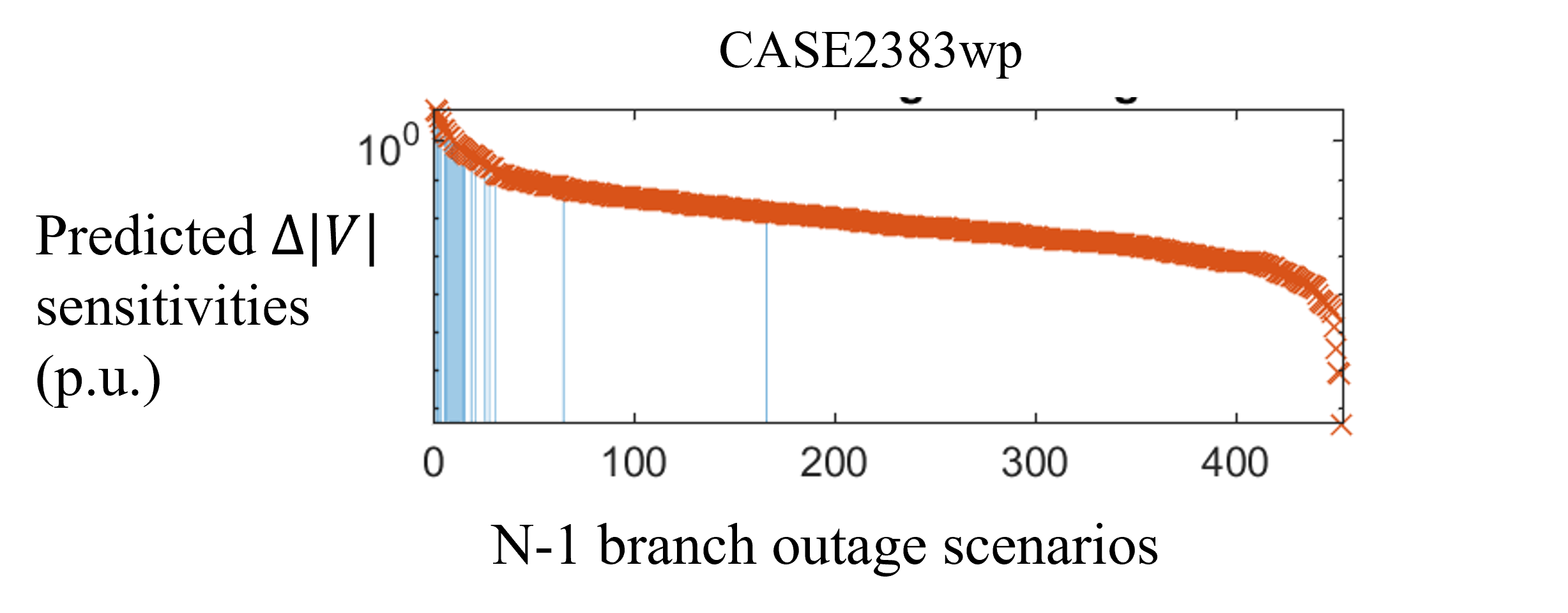}
	\caption[]{The proposed sensitivity gives larger values to critical contingencies: Red dots are $\Delta|V|$ sensitivities sorted in descending order; The blue lines marks branch outages that leads to islanding or blackout.
}
	\label{fig:LODF}
\end{figure}

Figure \ref{fig:case2383} shows an experiment on CASE2383wp where circuit-theoretic sensitivities for all possible N-1 line outages are calculated, to quantify the impact of each line outage event on  $|V|$ of the system. These predictions are then sorted in descending order for filtering out top $k$ serious outage events that the operator should pay more attention to.

\section{Conclusion}
This work proposes a AC sensitivity factor for line outage analysis, based on circuit theoretic approaches. Experiment results validated its efficacy in quantifying the impact of line outage, and its potential for fast contingency screening. 

\bibliographystyle{IEEEtran}
\bibliography{mybib}
\end{document}